\documentstyle[12pt,graphicx,amsfonts]{article}

\newcommand{\beq}{\begin{equation}}
\newcommand{\eeq}{\end{equation}}
\newcommand{\bea}{\begin{eqnarray}}
\newcommand{\eea}{\end{eqnarray}}
\newcommand{\ena}{\end{eqnarray}}

\def\mathfrak{\bf}

\renewcommand{\(}{\left(}
\renewcommand{\)}{\right)}
\renewcommand{\[}{\left[}
\renewcommand{\]}{\right]}

\def\be{\begin{equation}}
\def\ee{\end{equation}}
\def\bea{\begin{eqnarray}}
\def\eea{\end{eqnarray}}

\def\dt#1{\on{\hbox{\bf .}}{#1}}                
\def\Dot#1{\dt{#1}}
\def\IR{\relax{\rm I\kern-.18em R}}
\def\binomial#1#2{\left(\,{\buildrel
{\raise4pt\hbox{$\displaystyle{#1}$}}\over
{\raise-6pt\hbox{$\displaystyle{#2}$}}}\,\right)}

\def\[{\lfloor{\hskip 0.35pt}\!\!\!\lceil}
\def\]{\rfloor{\hskip 0.35pt}\!\!\!\rceil}

\newcommand{\AmS}{{\protect\the\textfont2
  A\kern-.1667em\lower.5ex\hbox{M}\kern-.125emS}}

\catcode`@=11
\def\un#1{\relax\ifmmode\@@underline#1\else
        $\@@underline{\hbox{#1}}$\relax\fi}
\catcode`@=12

\def\fracm#1#2{\hbox{\large{${\frac{{#1}}{{#2}}}$}}}

\def\ad{{\kern0.5pt
                   \alpha \kern-5.05pt
\raise5.8pt\hbox{$\textstyle.$}\kern
0.5pt}}

\def\Dot#1{{\kern0.5pt
     {#1} \kern-5.05pt \raise5.8pt\hbox{$\textstyle.$}\kern
0.5pt}}

\def\t{\tau}

\def\z{\zeta}

\def\O{\Omega}

\def\ca{{\cal A}}
\def\cb{{\cal B}}
\def\cc{{P}}

\def\cm{{\cal M}}
\def\cn{{\cal N}}

\def\cp{{\cal P}}
\def\cq{{\cal Q}}

\def\cs{{\cal S}}

\def\rI{{\rm I}}
\def\rJ{{\rm J}}
\def\rK{{\rm K}}
\def\rL{{\rm L}}
\def\rIJ{{\rm {I \, J}}}
\def\rKL{{\rm {K \, L}}}
\def\rIK{{\rm {I \, K}}}
\def\rIL{{\rm {I \, L}}}
\def\rJK{{\rm {J \, K}}}
\def\rJL{{\rm {J \, L}}}
\def\rIJKL{{\rm {IJKL}}}

\def\cnd{1D $\cn = 4$ \,}
\def\sg{{\Sf g}}
\def\sG{{\Sf G}}
\def\sh{{\Sf h}}

\def\bo{{\raise.15ex\hbox{\large$\Box$}}}               
\def\pa{\partial}                                       
\def\TH{{\raise.2ex\hbox{$\displaystyle \bigodot$}\mskip-4.7mu \llap H
\;}}
\def\face{{\raise.2ex\hbox{$\displaystyle \bigodot$}\mskip-2.2mu \llap
{$\ddot
        \smile$}}}                                      

\def\VEV#1{\left\langle #1\right\rangle}        
\def\abs#1{\left| #1\right|}                    
\def\leftrightarrowfill{$\mathsurround=0pt \mathord\leftarrow \mkern-6mu
        \cleaders\hbox{$\mkern-2mu \mathord- \mkern-2mu$}\hfill
        \mkern-6mu \mathord\rightarrow$}
\def\dvec#1{\vbox{\ialign{##\crcr
        \leftrightarrowfill\crcr\noalign{\kern-1pt\nointerlineskip}
        $\hfil\displaystyle{#1}\hfil$\crcr}}}           
\def\dt#1{{\buildrel {\hbox{\LARGE .}} \over {#1}}}     

\def\fracm#1#2{\hbox{\large{${\frac{{#1}}{{#2}}}$}}}
\def\frac#1#2{{\textstyle{#1\over\vphantom2\smash{\raise.20ex
        \hbox{$\scriptstyle{#2}$}}}}}                   
\def\sfrac#1#2{{\vphantom1\smash{\lower.5ex\hbox{\small$#1$}}\over
        \vphantom1\smash{\raise.4ex\hbox{\small$#2$}}}} 
\def\bfrac#1#2{{\vphantom1\smash{\lower.5ex\hbox{$#1$}}\over
        \vphantom1\smash{\raise.3ex\hbox{$#2$}}}}       
\def\afrac#1#2{{\vphantom1\smash{\lower.5ex\hbox{$#1$}}\over#2}}    
\def\on#1#2{\mathop{\null#2}\limits^{#1}}               

\newskip\humongous \humongous=0pt plus 1000pt minus 1000pt

\newif\ifdtup

  \def\pp{{\mathchoice
          {
              \kern 1pt%
              \raise 1pt
              \vbox{\hrule width5pt height0.4pt depth0pt
                    \kern -2pt
                    \hbox{\kern 2.3pt
                          \vrule width0.4pt height6pt depth0pt
                          }
                    \kern -2pt
                    \hrule width5pt height0.4pt depth0pt}%
                    \kern 1pt
           }
            {
              \kern 1pt%
              \raise 1pt
              \vbox{\hrule width4.3pt height0.4pt depth0pt
                    \kern -1.8pt
                    \hbox{\kern 1.95pt
                          \vrule width0.4pt height5.4pt depth0pt
                          }
                    \kern -1.8pt
                    \hrule width4.3pt height0.4pt depth0pt}%
                    \kern 1pt
            }
            {
              \kern 0.5pt%
              \raise 1pt
              \vbox{\hrule width4.0pt height0.3pt depth0pt
                    \kern -1.9pt  
                    \hbox{\kern 1.85pt
                          \vrule width0.3pt height5.7pt depth0pt
                          }
                    \kern -1.9pt
                    \hrule width4.0pt height0.3pt depth0pt}%
                    \kern 0.5pt
            }
            {
              \kern 0.5pt%
              \raise 1pt
              \vbox{\hrule width3.6pt height0.3pt depth0pt
                    \kern -1.5pt
                    \hbox{\kern 1.65pt
                          \vrule width0.3pt height4.5pt depth0pt
                          }
                    \kern -1.5pt
                    \hrule width3.6pt height0.3pt depth0pt}%
                    \kern 0.5pt
            }
        }}

  \def\mm{{\mathchoice
                       {
                             \kern 1pt
               \raise 1pt    \vbox{\hrule width5pt height0.4pt depth0pt
                                  \kern 2pt
                                  \hrule width5pt height0.4pt depth0pt}
                             \kern 1pt}
                       {
                            \kern 1pt
               \raise 1pt \vbox{\hrule width4.3pt height0.4pt depth0pt
                                  \kern 1.8pt
                                  \hrule width4.3pt height0.4pt depth0pt}
                             \kern 1pt}
                       {
                            \kern 0.5pt
               \raise 1pt
                            \vbox{\hrule width4.0pt height0.3pt depth0pt
                                  \kern 1.9pt
                                  \hrule width4.0pt height0.3pt depth0pt}
                            \kern 1pt}
                       {
                           \kern 0.5pt
             \raise 1pt  \vbox{\hrule width3.6pt height0.3pt depth0pt
                                  \kern 1.5pt
                                  \hrule width3.6pt height0.3pt depth0pt}
                           \kern 0.5pt}
                       }}

\def\pd{{\kern0.5pt
                   + \kern-5.05pt \raise5.8pt\hbox{$\textstyle.$}\kern
0.5pt}}

\def\pmd{{\kern0.5pt
                  \pm \kern-5.05pt \raise6.3pt\hbox{$\textstyle.$}\kern1.5pt}}

\def\md{{\mathchoice
   {
      {{\kern 1pt - \kern-6.2pt \raise5pt\hbox{$\textstyle.$}\kern 1pt}}}
    {
      {{\kern 1pt - \kern-6.2pt \raise5pt\hbox{$\textstyle.$}\kern 1pt}}}
    {
      {\kern0.5pt - \kern-5.05pt \raise3.4pt\hbox{$\textstyle.$}\kern0.5pt}}
    {
      {\kern0.5pt - \kern-5.05pt \raise3.4pt\hbox{$\textstyle.$}\kern0.5pt}}}}

\def\ad{{\dot{\alpha}}}

\def\pp{{\mathchoice
          {
              \kern 1pt%
              \raise 1pt
              \vbox{\hrule width5pt height0.4pt depth0pt
                    \kern -2pt
                    \hbox{\kern 2.3pt
                          \vrule width0.4pt height6pt depth0pt
                          }
                    \kern -2pt
                    \hrule width5pt height0.4pt depth0pt}%
                    \kern 1pt
           }
            {
              \kern 1pt%
              \raise 1pt
              \vbox{\hrule width4.3pt height0.4pt depth0pt
                    \kern -1.8pt
                    \hbox{\kern 1.95pt
                          \vrule width0.4pt height5.4pt depth0pt
                          }
                    \kern -1.8pt
                    \hrule width4.3pt height0.4pt depth0pt}%
                    \kern 1pt
            }
            {
              \kern 0.5pt%
              \raise 1pt
              \vbox{\hrule width4.0pt height0.3pt depth0pt
                    \kern -1.9pt  
                    \hbox{\kern 1.85pt
                          \vrule width0.3pt height5.7pt depth0pt
                          }
                    \kern -1.9pt
                    \hrule width4.0pt height0.3pt depth0pt}%
                    \kern 0.5pt
            }
            {
              \kern 0.5pt%
              \raise 1pt
              \vbox{\hrule width3.6pt height0.3pt depth0pt
                    \kern -1.5pt
                    \hbox{\kern 1.65pt
                          \vrule width0.3pt height4.5pt depth0pt
                          }
                    \kern -1.5pt
                    \hrule width3.6pt height0.3pt depth0pt}%
                    \kern 0.5pt
            }
        }}

  \def\mm{{\mathchoice
                       {
                             \kern 1pt
               \raise 1pt    \vbox{\hrule width5pt height0.4pt depth0pt
                                  \kern 2pt
                                  \hrule width5pt height0.4pt depth0pt}
                             \kern 1pt}
                       {
                            \kern 1pt
               \raise 1pt \vbox{\hrule width4.3pt height0.4pt depth0pt
                                  \kern 1.8pt
                                  \hrule width4.3pt height0.4pt depth0pt}
                             \kern 1pt}
                       {
                            \kern 0.5pt
               \raise 1pt
                            \vbox{\hrule width4.0pt height0.3pt depth0pt
                                  \kern 1.9pt
                                  \hrule width4.0pt height0.3pt depth0pt}
                            \kern 1pt}
                       {
                           \kern 0.5pt
             \raise 1pt  \vbox{\hrule width3.6pt height0.3pt depth0pt
                                  \kern 1.5pt
                                  \hrule width3.6pt height0.3pt depth0pt}
                           \kern 0.5pt}
                       }}

\def\pd{{\kern0.5pt
                   + \kern-5.05pt \raise5.8pt\hbox{$\textstyle.$}\kern
0.5pt}}

\def\pmd{{\kern0.5pt
                  \pm \kern-5.05pt \raise6.3pt\hbox{$\textstyle.$}\kern1.5pt}}

\def\md{{\mathchoice
   {
      {{\kern 1pt - \kern-6.2pt \raise5pt\hbox{$\textstyle.$}\kern 1pt}}}
    {
      {{\kern 1pt - \kern-6.2pt \raise5pt\hbox{$\textstyle.$}\kern 1pt}}}
    {
      {\kern0.5pt - \kern-5.05pt \raise3.4pt\hbox{$\textstyle.$}\kern0.5pt}}
    {
      {\kern0.5pt - \kern-5.05pt \raise3.4pt\hbox{$\textstyle.$}\kern0.5pt}}}}

\def\dslash{\not{\hbox{\kern-2pt $\partial$}}}
\def\Dslash{\not{\hbox{\kern-4pt $D$}}}
\def\pslash{\not{\hbox{\kern-2.3pt $p$}}}
 \newtoks\slashfraction
 \slashfraction={.13}
 \def\slash#1{\setbox0\hbox{$ #1 $}
 \setbox0\hbox to \the\slashfraction\wd0{\hss \box0}/\box0 }

\def\Sf#1{{\hbox{\sf #1}}}      
\font\ro=cmsy10                          
\def\kcr{{\hbox{\ro \char'170}}}                
\def\ktl{{\hbox{\ro \char'170}}}        
\def\ktr{{\hbox{\ro \char'170}}}        
\def\kbl{{\hbox{\ro \char'170}}}        
\def\kbr{{\hbox{\ro \char'170}}}        

\def\plpl{\raise-2pt\hbox{$\raise3pt\hbox{$_+$}\hskip-6.67pt\raise0.0pt
\hbox{$^+$}\hskip 0.01pt$}}
\def\mimi{\raise-2pt\hbox{$\raise3pt\hbox{$_-$}\hskip-6.67pt\raise0.0pt
\hbox{$^-$}\hskip 0.01pt$}}

\def\bo{{\raise.15ex\hbox{\large$\Box$}}}               
\def\pa{\partial}                                       
\def\TH{{\raise.2ex\hbox{$\displaystyle \bigodot$}\mskip-4.7mu \llap H \;}}
\def\face{{\raise.2ex\hbox{$\displaystyle \bigodot$}\mskip-2.2mu \llap {$\ddot
        \smile$}}}                                      

\def\VEV#1{\left\langle #1\right\rangle}        
\def\abs#1{\left| #1\right|}                    
\def\leftrightarrowfill{$\mathsurround=0pt \mathord\leftarrow \mkern-6mu
        \cleaders\hbox{$\mkern-2mu \mathord- \mkern-2mu$}\hfill
        \mkern-6mu \mathord\rightarrow$}
\def\dvec#1{\vbox{\ialign{##\crcr
        \leftrightarrowfill\crcr\noalign{\kern-1pt\nointerlineskip}
        $\hfil\displaystyle{#1}\hfil$\crcr}}}           
\def\dt#1{{\buildrel {\hbox{\LARGE .}} \over {#1}}}     

\def\fracm#1#2{\hbox{\large{${\frac{{#1}}{{#2}}}$}}}
\def\frac#1#2{{\textstyle{#1\over\vphantom2\smash{\raise.20ex
        \hbox{$\scriptstyle{#2}$}}}}}                   
\def\sfrac#1#2{{\vphantom1\smash{\lower.5ex\hbox{\small$#1$}}\over
        \vphantom1\smash{\raise.4ex\hbox{\small$#2$}}}} 
\def\bfrac#1#2{{\vphantom1\smash{\lower.5ex\hbox{$#1$}}\over
        \vphantom1\smash{\raise.3ex\hbox{$#2$}}}}       
\def\afrac#1#2{{\vphantom1\smash{\lower.5ex\hbox{$#1$}}\over#2}}    
\def\on#1#2{\mathop{\null#2}\limits^{#1}}               

\parskip=\medskipamount                 
\thicklines                         

\def\oldheadpic{                                
        \setlength{\unitlength}{.4mm}
        \thinlines
        \par
        \begin{picture}(349,16)
        \put(325,16){\line(1,0){4}}
        \put(330,16){\line(1,0){4}}
        \put(340,16){\line(1,0){4}}
        \put(335,0){\line(1,0){4}}
        \put(340,0){\line(1,0){4}}
        \put(345,0){\line(1,0){4}}
        \put(329,0){\line(0,1){16}}
        \put(330,0){\line(0,1){16}}
        \put(339,0){\line(0,1){16}}
        \put(340,0){\line(0,1){16}}
        \put(344,0){\line(0,1){16}}
        \put(345,0){\line(0,1){16}}
        \put(329,16){\oval(8,32)[bl]}
        \put(330,16){\oval(8,32)[br]}
        \put(339,0){\oval(8,32)[tl]}
        \put(345,0){\oval(8,32)[tr]}
        \end{picture}
        \par
        \thicklines
        \vskip.2in}
\def\oldtitle#1#2#3#4{\oldheadpic\begin{center}\vglue.5in{\large\bf #1}\\[.6in]
        {#2}\\[.1in] {\it Department of Physics and Astronomy}\\
        {\it University of Maryland, College Park, MD 20742}\\[.6in]
        Physics Publication \#{#3}\\ {#4}\\[1.5in] {\bf ABSTRACT}\\[.1in]
        \end{center} \begin{quotation}}                 
\def\oldTitle#1#2#3#4#5#6#7{\oldheadpic\begin{center} \vglue .4in
        {\large\bf #1}\\[.4in]
        {#2}\\[.1in] {\it Department of Physics and Astronomy}\\
        {\it University of Maryland, College Park, MD 20742}\\[.1in]
        {#3}\\[.1in] {\it {#4}}\\ {\it {#5}}\\[.4in]
        Physics Publication \#{#6}\\ {#7}\\[.5in] {\bf ABSTRACT}\\[.1in]
        \end{center} \begin{quotation}}                 
\def\border{                                            
        \setlength{\unitlength}{1mm}
        \newcount\xco
        \newcount\yco
        \xco=-21
        \yco=12
        \begin{picture}(140,0)
        \put(\xco,\yco){$\ktl$}
        \advance\yco by-1
        {\loop
        \put(\xco,\yco){$\kcr$}
        \advance\yco by-2
        \ifnum\yco>-240
        \repeat
        \put(\xco,\yco){$\kbl$}}
        \xco=158
        \yco=12
        \put(\xco,\yco){$\ktr$}
        \advance\yco by-1
        {\loop
        \put(\xco,\yco){$\kcr$}
        \advance\yco by-2
        \ifnum\yco>-240
        \repeat
        \put(\xco,\yco){$\kbr$}}
        \put(-20,13){\tiny **University of Maryland * Center for String and
         Particle  Theory* Physics Department***University of Maryland *Center
        for String and Particle  Theory** }
        \put(-20,-241.5){\tiny **University of Maryland * Center for String and
         Particle  Theory* Physics Department***University of Maryland *Center
        for String and Particle  Theory** }
        \end{picture}
        \par\vskip-8mm}
\def\bordero{                                           
        \setlength{\unitlength}{1mm}
        \newcount\xco
        \newcount\yco
        \xco=-31
        \yco=12
        \begin{picture}(140,0)
        \put(\xco,\yco){$\ktl$}
        \advance\yco by-1
        {\loop
        \put(\xco,\yco){$\kclr$}
        \advance\yco by-2
        \ifnum\yco>-240
        \repeat
        \put(\xco,\yco){$\kbl$}}
        \xco=151
        \yco=12
        \put(\xco,\yco){$\ktr$}
        \advance\yco by-1
        {\loop
        \put(\xco,\yco){$\kcr$}
        \advance\yco by-2
        \ifnum\yco>-240
        \repeat
        \put(\xco,\yco){$\kbr$}}
        \put(-20,12){\ooo bacdefghidfghghdhededbihdgdfdfhhdheidhdhebaaahjhhdahba

hgdedge
   hgfdiehhgdigicba}
        \put(-20,-241.5){\ooo ababaighefdbfghgeahgdfgafagihdidihiidhiagfedhadbfd

ecdcdfa
   gdcbhaddhbgfchbgfdacfediacbabab}
        \end{picture}
        \par\vskip-8mm}
\def\headpic{                                           
        \indent
        \setlength{\unitlength}{.4mm}
        \thinlines
        \par
        \begin{picture}(29,16)
        \put(165,16){\line(1,0){4}}
        \put(170,16){\line(1,0){4}}
        \put(180,16){\line(1,0){4}}
        \put(175,0){\line(1,0){4}}
        \put(180,0){\line(1,0){4}}
        \put(185,0){\line(1,0){4}}
        \put(169,0){\line(0,1){16}}
        \put(170,0){\line(0,1){16}}
        \put(179,0){\line(0,1){16}}
        \put(180,0){\line(0,1){16}}
        \put(184,0){\line(0,1){16}}
        \put(185,0){\line(0,1){16}}
        \put(169,16){\oval(8,32)[bl]}
        \put(170,16){\oval(8,32)[br]}
        \put(179,0){\oval(8,32)[tl]}
        \put(185,0){\oval(8,32)[tr]}
        \end{picture}
        \par\vskip-6.5mm
        \thicklines}
\def\title#1#2#3#4{\border\headpic {\hbox to\hsize{#4 \hfill UMDEPP #3}}\par
        \begin{center} \vglue .5in {\large\bf #1}\\[.6in]
        {#2}\\[.1in] {\it Department of Physics and Astronomy}\\
        {\it University of Maryland, College Park, MD 20742}\\[1.5in]
        {\bf ABSTRACT}\\[.1in] \end{center} \begin{quotation}}  
\def\Title#1#2#3#4#5#6#7{\border\headpic
        {\hbox to\hsize{#7 \hfill UMDEPP #6}}\par
        \begin{center} \vglue .4in {\large\bf #1}\\[.4in]
        {#2}\\[.1in] {\it Department of Physics and Astronomy}\\
        {\it University of Maryland, College Park, MD 20742}\\[.1in]
        {#3}\\[.1in] {\it {#4}}\\ {\it {#5}}\\[.5in] {\bf ABSTRACT}\\[.1in]
        \end{center} \begin{quotation}}                 
\def\endtitle{\end{quotation}\newpage}                  

\def\qd{{\kern0.5pt
                   q \kern-5.05pt \raise5.8pt\hbox{$\textstyle.$}\kern
0.5pt}}

\begin{document}

\def\dt#1{\on{\hbox{\bf .}}{#1}}                
\def\Dot#1{\dt{#1}}

\def\gfrac#1#2{\frac {\scriptstyle{#1}}
        {\mbox{\raisebox{-.6ex}{$\scriptstyle{#2}$}}}}
\def\gg{{\hbox{\sc g}}}
\border\headpic {\hbox to\hsize{January 2008 \hfill
{UMDEPP 07-016}}}
\par
{$~$ \hfill
{~}}
\par

\setlength{\oddsidemargin}{0.3in}
\setlength{\evensidemargin}{-0.3in}
\begin{center}
\vglue .10in
{\large\bf Short Distance Operator Product Expansion of the\\ 1D, ${\cal N}$ = 4
Extended ${\cal GR}$ Super Virasoro Algebra by \\ Use of Coadjoint
Representations}\\[.5in]

Isaac Chappell II\footnote{ichappel@umd.edu}  and 
S.\, James Gates, Jr.\footnote{gatess@wam.umd.edu}\\[0.3in]
${}^\dag${\it Center for String and Particle Theory\\
Department of Physics, University of Maryland\\
College Park, MD 20742-4111 USA}\\[1.8in]

{\bf ABSTRACT}\\[.01in]
\end{center}
\begin{quotation}
{Using the previous construction of the geometrical representation (${\cal GR}$) 
of the centerless 1D, $\cal N$ = 4 extended Super Virasoro algebra, we construct 
the corresponding Short Distance Operation Product Expansions for the deformed
version of the algebra.  This algebra differs from the regular algebra by the addition 
of terms containing the Levi{}-Civita tensor.  How this addition changes the 
super{}-commutation relations and affects the Short Distance Operation Product 
Expansions (OPEs) of the associated fields is investigated.  The Method of Coadjoint 
Orbits, which removes the need first to find Lagrangians invariant under the action
of the symmetries, is used to calculate the expansions.  Finally, an alternative 
method involving Clifford algebras is investigated for comparison.}

\endtitle

\setlength{\oddsidemargin}{0.3in}
\setlength{\evensidemargin}{-0.3in}

\setcounter{equation}{0}
\section{Introduction}

$~~~$ One of the fundamental mathematical objects of String Theory (ST) 
is the Virasoro algebra. \ It is used in the description of simple open/closed
strings and is well{}-developed in Conformal Field Theory (CFT), a primary 
tool for probing strings.   A familiar technique from CFT commonly used in this 
context is the Operator Product Expansion (OPE) as it is closely related to 
the calculation of two{}-point correlation functions which themselves
are related to the propagation and interaction of fields represented in ST.

In many discussions, the beginning of such constructions involves first
finding an action (containing appropriate fields) that is invariant under a 
realization of the (super)conformal symmetry group.  The solutions of the
fields equations of motion are expanded in terms of Fourier series.  The 
Noether charges associated with the generators are, using their expressions 
in terms of the fields, also then expressed in terms of such Fourier series. 
Finally OPE's are then calculated.  Clearly the role of the action is prominent,
both in determing the Noether Charges and the field equations of
motion.

Instead the method to be used in this work for calculating these OPE's is the 
Coadjoint Orbit method developed by Kirillov \cite{Kirillov} and built upon the 
elements of Lie algebras and their realizations.  One goes from the closed 
algebra of operators to elements of a vector space.  This vector space is then 
expanded by the addition of a dual space of covectors and a bilinear metric 
between the two.  These objects are then used to find the coadjoint orbits which  
can be used in the OPE.

In this paper, the algebra used is the extended ${\cal GR}$ Super Virasoro
Algebra\footnote{It is perhaps more accurate to describe out starting point
as a 'Witt algebra' as it is 'centerless' $~~~~~~$ and constructed from the vector fields
associated with superspace coordinates. } (${\cal GR}$ SVA), a much larger 
algebra than the one related to the Virasoro algebra. The Virasoro algebra 
in this case contains familiar time{}-space operators from the Poincare algebra 
and the conformal algebra. The extended ${\cal GR}$ Super Virasoro algebra 
has many parts.  It is the Super Virasoro algebra because of the inclusion of 
supersymmetric elements and thus an enlarged symmetry.  ${\cal GR}$ stands 
for ``Geometrical Representation'' which means that methods involve groups and algebras
acting on the operators described in a representation as derivations on
the superspace.  It is extended because it can be considered on the basis of
the coordinates of  $\cal N$-extended superspace.  In the special case of $\cal 
N$ = 4, there is a deformation due to the addition of certain terms in the derivations 
as a function of a variable $\ell$. These terms change the number of operators 
in the algebra for certain values of $\ell$.

The calculations to be done in this work are in the context of one temporal 
dimension (1D) and four fermionic or Grassmann dimensions ($ {\cal N}$ = 4).    
There is a relationship between 4D, $\cal N$ = 1 theories and 1D, $\cal N$ = 4 
theories but that will not be discussed in this paper.

The outline of the paper is as follows.  The first section will describe the 
extended ${\cal GR}$ Super Virasoro algebra with a focus on under what
conditions does it close and the corresponding number of operators.  The 
next section will explain how the Coadjoint Orbit method is used and its 
application on the algebra. The third section brings in the OPE and it will 
be used to calculate various short distance Operator Product Expansions 
for the algebra.  In Section \ref{clifford}, a different look at the whole method 
from the use of Clifford algebras as an alternative to derivations.  Finally,
there will be a discussion of some of the implications of the results of the 
paper.

\setcounter{equation}{0}
\section{Realization of the 1D, $\cal N$ = 4 Extended GR \\ Super Virasoro
Algebra}

$~~~$ One can get to the Super Virasoro algebra by starting with the
SO($\cal N$) algebra using $T_{IJ }$ as generators.  Adding translations 
generated by momentum generators $P$, the dilations, ${\Delta}$, 
special conformal transformations, $K$, the supersymmetry generators
$Q_{\rI}$, and finally and $S_{\rI}$, make up the superconformal 
algebra. The total algebra can be represented as derivations with respect
to the superspace.  It has some peculiar properties that still remain a
mysterious.   For $\cal N$ $\le$ 4, it is known how to close this algebra
without additional generators.  It is clear however, that for $\cal N$ $>$ 4
closure requires the presence of additional operators.

In any event, the operators can be represented by derivations of the one dimensional 
time variable and its derivative, $\t$ and ${\partial}{}_{\t}$, and the $\cn$ = 4
superspace variables and their derivatives, $\zeta{}^{\rI}$ and ${\partial}{}_{
\rI}$.  The time variable and its derivative are real and commute with everything.
The superspace coordinates are real Grassmann variables, anti{}-commuting (
${\zeta}{}^{I }$ ${\zeta}{}^{J }$ = $-$  ${\zeta}{}^{J }$ ${\zeta}{}^{I }$ ) 
and squaring to zero ( [${\zeta}{}^{\rI}]^2  = 0$).  The algebra is defined by its commutation relations.  There are 36
possible combinations but only thirteen are nonzero:
\begin{eqnarray}
\left\[~ \Delta \, , \, P \} \right. &=& -i P ~~~,~~~
\left\[~ \Delta \, , \, Q_{\rI} \} \right. \,=\, -i \fracm{1}{2} \, Q_{\rI} ~~~,~~~
\left\[~ \Delta \, , \, K \} \right. \,=\, iK ~~~,  \\
\left\[~ \Delta \, , \, S_{\rI} \} \right.&=& i \, \fracm 12 \, S_{\rI} ~~~,~~~
\left\[~ P \, , \, S_{\rI} \} \right.\,=\,  i Q_{\rI} ~~~,~~~
\left\[~ K \, , \, Q_{\rI} \} \right.\,=\, -i S_{\rI} ~~~,~~~\\
\left\[~ Q_{\rI} \, , \, Q_{\rJ} \} \right.&=& 4 \delta_{\rIJ} \, P ~~~,~~~
\left\[~ S_{\rI} \, , \, S_{\rJ} \} \right.\,=\, 4 \delta_{\rIJ} K ~~~, ~~~
\left\[~ P \, , \, K\} \right. \,=\, -i 2 \Delta ~~~, \\
\left\[~ Q_{\rI} \, , \, S_{\rJ} \} \right. &=& 4 \, \delta_{\rIJ} \,\Delta + 2T_{
\rIJ} ~~~, \\
\left\[~ T_{\rIJ} \, , \, Q_{\rK} \} \right. &=& -i \delta_{\rIK} Q_{\rJ} + i \delta_{
\rJK} \, Q_{\rI} ~~~, \\
\left\[~ T_{\rIJ} \, , \, S_{\rK} \} \right. &=& -i \delta_{\rIK} \, S_{\rJ}+ i \delta_{
\rJK} \, S_{\rI} ~~~, \\
\left\[~ T_{\rIJ} \, , \, T_{\rKL} \} \right. &=& i\delta_{\rJK} \, T_{\rIL} -i \delta_{
\rJL} \, T_{\rIK} + i \delta_{\rIL} \, T_{\rJK} + i\delta_{\rIK} \, T_{\rJL} 
\end{eqnarray}

The generators and their 
corresponding symmetries are listed in Table I.

\begin{center}
\begin{tabular}{|c|c|c|c|}
\hline
\multicolumn{4}{|c|}{ \bf Table 1: Generators and Their Associated Symmetries and Derivations} 
\\ \hline\hline
${\rm Generators}$ & ${\rm Symmetry}$ & ${\rm Derivation}$ & ${\rm {No.\, of\, generators}}$ \\ \hline
$ P $ & Translations & $ i \,\pa_{\t} $ & 1 \\ \hline
$ {\Delta} $ & Dilations & $ i ( \t \pa_{\t} + \fracm 12  \z^{\rI} \pa_{\rI} )$ & 1 \\ \hline
$ K $ & Special Conformal & $ i ( \t^{2} \pa_{\t} +  \t \z^{\rI} \pa_{\rI} )$  & 1 \\ \hline
$ Q_{\rI } $ & Supersymmetry & $ i (\,  \pa_{\rI} - i \, 2 \, \z_{\rI} \pa_{\t} ) $ & $4 = [\cn ]$ \\ \hline
$ S_{\rI } $ & S-supersymmetry & $ i \t \pa_{\rI} + 2 \t \z_{\rI} \pa_{\t} + \z_{\rI} \z^{\rJ} \pa_{\rJ} $ & $4 = [\cn ]$ \\ \hline
$ T_{\rIJ} $ & SO($\cal N$) & $ i (\z_{\rI} \, \pa_{\rJ} - \z_{\rJ} \, \pa_{\rI} )$ & $  6 = [{\cal N} ({\cal N} -1)/2] $ \\ \hline
\end{tabular}
\end{center}
\vskip.2in
\centerline{{\bf Table 1}} 

This algebra can be deformed in $\cn =4$ with the addition of a Levi{}-Civita
tensor, ${\epsilon}_{\rIJKL}$, and a parameter, $\ell$, that measures the deformation. 
\ It only affects three of the six operators:
\begin{eqnarray}
S_{\rI}(\ell) &\equiv& i \tau \partial_{\tau } + 2 \tau \zeta_{\rI} \partial_{\tau } + 
2 \zeta_{\rI} \zeta^{\rJ} \partial_{\rJ} + \ell \epsilon_{\rIJKL} (\zeta^{\rJ} \zeta^{
\rK} \partial^{\rL} - \frac{1}{3!} \zeta^{\rJ} \zeta^{\rK} \zeta^{\rL} \partial_{\tau }) \\
K(\ell) & \equiv & i ( \tau^{2} \partial_{\tau } + \tau \zeta^{\rI} \partial_{\rI} -i \, 2 
\, \ell \epsilon^{\rIJKL} [ \fracm{1}{4} \zeta_{\rI} \zeta_{\rJ} \zeta_{\rK} \partial_{
\rL} + \zeta_{\rI} \zeta_{\rJ} \zeta_{\rK} \zeta_{\rL} \, \pa_{\t}  \, ]  )  \\
T_{\rIJ}(\ell) & \equiv & i \zeta_{[\rI} \partial_{\rJ]} -i \ell \epsilon_{\rIJKL} \zeta_{
\rK} \partial_{\rL} 
\end{eqnarray}

This changes the last three of the commutation relations
\begin{eqnarray}
\left\[~ T_{\rIJ} \, , \, Q_{\rK} \} \right. &=& -i \delta_{\rIK} Q_{\rJ} + i \delta_{\rJK} 
\, Q_{\rI} + i \, \ell \, \epsilon_{\rIJKL} Q_{\rL} \\
\left\[~ T_{\rIJ} \, , \, S_{\rK} \} \right. &=& -i \delta_{\rIK} \, S_{\rJ}+ i \delta_{\rJK} 
\, S_{\rI} + i \, \ell \, \epsilon_{\rIJKL} S_{\rL}~~~, \\
\left\[~ T_{\rIJ} \, , \, T_{\rKL} \} \right. &=& \fracm 12 \(\ell^2 +3\) \[i\delta_{\rJK} 
\, T_{\rIL} -i \delta_{\rJL} \, T_{\rIK} + i \delta_{\rIL} \, T_{\rJK} + i\delta_{\rIK} 
\, T_{\rJL} \] \nonumber \\
&+& \fracm 12 \(\ell^2 - 1\) \[i\delta_{\rJK} \, Y_{\rIL} -i \delta_{\rJL} \, Y_{\rIK} 
+ i \delta_{\rIL} \, Y_{\rJK} + i\delta_{\rIK} \, Y_{\rJL} \] ~~~  
\end{eqnarray}
with $
Y_{\rIJ} \equiv i\zeta_{[\rI} \partial_{\rJ]} \, + \, i \, \ell \, \epsilon_{\rIJKL} \zeta_{
\rK} \partial_{\rL}$.  For $\ell\, \, \pm =1$, there are no $Y_{\rIJ}$ terms in the last 
commutation relation.

The next step is to recast the previous generators in form in which the
relationship to the super Virasoro algebra is more obvious. \ This is
done by choosing the forms
\begin{eqnarray}
L_{m} \equiv  -[\tau^{m+1} \partial_{\tau } + \frac{1}{2} (m+1)\tau^{m} \zeta 
\partial _{\zeta }] & , &
H_{r} \equiv  -[\tau^{r+1} \partial_{\tau } + \frac{1}{2} (r+1)\tau^{r} \zeta 
\partial _{\zeta }] \\
F_{m} \equiv  i\tau^{m+\frac{1}{2}} [\partial_{\zeta } \, - i \, 2 \zeta \partial_{\tau 
}] & , & 
G_{r} \equiv  i\tau^{r+\frac{1}{2}} [\partial_{\zeta } - \, i \, 2 \zeta \partial_{\tau }]  
\end{eqnarray}
where  $m\in \mathbb{Z}$ and  $r\in \mathbb{Z} + \fracm 12 $.  The L and H 
are the same except L takes integers and H takes half integers.  The F and 
G forms follow the same pattern.  H is fermionic and L is bosonic because L 
exists in the $\cn=0$ case.  If one looks at the lowest level  of the set of (L, 
H, F, G) generators, some of the previous generators are now represented: 

\begin{equation}
P\rightarrow L_{-1},\Delta \rightarrow L_{0},K\rightarrow
L_{+1},Q\rightarrow G_{-{\frac{1}{2}}},S\rightarrow G_{+{\frac{1}{2}}}
\end{equation}
The $T_{\rIJ}$ generators remain the same.

 These new generator pairs can be combined using a different notation
with simple commutation relations:
\begin{equation}
\left(
\begin{array}{ccc}
L_{\ca} &\equiv & (L_{m},H_{r})\\
G_{\ca} &\equiv &(F_{m},G_{r})
\end{array} 
\right)
\to
\left(
\begin{array}{ccc}
\[L_{\ca}\, , \, L_{\cb}\} & = & (\ca-\cb ) L_{\ca + \cb} \\
\[G_{\ca}\, ,\, G_{\cb} \} & = & - i \, 4 \, L_{\ca + \cb} \\
\[L_{\ca}\, , \,G_{\cb} \} & = & (\frac{1}{2} \ca -\cb ) \, G_{\ca + \cb}
\end{array}
\right)
\end{equation}
with $\ca,\cb$ taking values in $\mathbb{Z}$ and $\mathbb{Z} + \fracm 12$. 
For $\cn =1$, this pair of generators is closed under graded commutation. In 
the \cnd exceptional Super Virasoro algebra, an index $I$ for the supersymmetric 
levels has to be added and the $\ell${}-deformed terms must be put in properly, 
including a $\ell${}-deformed supersymmetric $T_{\rIJ}(\ell)$ generator. For the 
\cnd exceptional Super Virasoro algebra, the set of generators ($L_{\ca}(\ell)$, 
$G_{\ca}^{\rI}(\ell)$, $T_{\ca}^{\rIJ }(\ell)$) closes under graded commutation.  
These generators are
\begin{eqnarray}
L_{\ca} &\equiv & -[\tau ^{\ca+1}\partial_{\tau }+\frac{1}{2} (\ca+1) \tau^{\ca} 
\zeta^{\rI} \partial_{\rI}] \cr
&+& i \ell \ca (\ca+1)\tau ^{\ca-1} [\zeta^{(3)\rI} \partial_{\rI} + i 4 \zeta^{(4)} 
\partial_{\tau}]\\
G_{\ca}^{\rI} &\equiv& \tau^{\ca+\frac{1}{2}}[\partial^{\rI} - i 2 \zeta^{\rI} 
\partial_{\tau }] + 2 ( \ca+\frac{1}{2}) \tau^{\ca-\frac{1}{2}} \zeta^{\rI} 
\zeta^{\rK}\partial _{\rK} \cr
&+& \ell (\ca+\frac{1}{2}) \tau^{\ca-\frac{1}{2}}[\epsilon^{\rIJKL} 
\zeta_{\rJ}\zeta _{\rK}\partial_{\rL} \cr
&-& i 4 \zeta^{(3) \rI} \partial_{\tau }] + i 4 \ell ({\ca}^{2}-\frac{1}{4})
\tau^{\ca-\frac{3}{2}} \zeta^{(4)} \partial^{\rI}\\
T_{\ca}^{{\rIJ}} &\equiv& \tau^{\ca}[\zeta^{[\rI}\partial^{\rJ]} - \ell \epsilon^{{
\rIJKL}} \zeta_{\rK} \partial_{\rL}]- i 2 \ell \ca   \tau^{\ca-1} [\zeta ^{(3)[\rI} 
\partial^{\rJ]}-\ell\epsilon^{{\rIJKL}} \zeta _{\rK}^{(3)} \partial_{\rL}]
\end{eqnarray}

Their supercommutation relations are
\begin{eqnarray}
\[L_{\ca} \, , \, L_{\cb}\]&=&(\ca - \cb) L_{\ca + \cb} + \frac{1}{8} \, c \, (\ca^{
3}-\ca) \delta_{\ca + \cb,0}\\
\[L_{\ca} \, , \,G_{\cb}^{\rI}\]&=&(\frac{\ca}{2}-\cb) G_{\ca + \cb}^{\rI}\\
\[L_{\ca} \, , \,T_{\cb}^{{\rIJ}}\]&=&-\cb T_{\ca + \cb}^{{\rIJ}}\\
\{G_{\ca}^{\rI} \, , \,G_{\cb}^{\rJ}\}&=&-i4\delta^{{\rIJ}} \, L_{\ca + \cb}-i \,2
 \,(\ca -\cb ) T_{\ca + \cb}^{{\rIJ}}-i\, c\,( \ca^{2}-\frac{1}{4})\delta_{\ca + 
 \cb,0}\delta ^{{\rIJ}}\\
\[T_{\ca}^{{\rIJ}} \, , \,G_{\cb}^{\rK}\]&=&\, 2 \,(\delta^{{\rJK}} \, G_{\ca + 
\cb}^{\rI}-\delta^{{\rIK}} \, G_{\ca + \cb}^{\rJ})\\
\[T_{\ca}^{{\rIJ}} \, , \,T_{\cb}^{{\rKL}}\]&=&T_{\ca + \cb}^{{\rIK}} \, \delta^{{
\rJL}}-\,T_{\ca + \cb}^{{\rIL}} \,\delta^{{\rJK}}+T_{\ca + \cb}^{{\rJL}} \, 
\delta^{{\rIK}} \, - \, T_{\ca + \cb}^{{\rJK}} \, \delta^{{\rIL}} \cr 
& & -2c(\ca - \cb) \,( \delta ^{\rI [\rK|}\delta ^{\rJ| \rL]})
\end{eqnarray}

A number of interesting points can be found here.  In previous papers 
\cite{Lubna} \cite{Curto}, the non{}-deformed ($\ell=0$) \cnd  ${\cal {GR}}$ 
Super Virasoro algebra is used to generate OPEs.  This algebra is the ``large'' 
$\cn = 4$ algebra which has a 16{}-dimensional representation.  It does 
not close unless two more {\it sets} of generators (U's and R's, which are 
related to the T's,) are added.  The $\ell=\pm 1$ cases of the $\ell${}-extended 
algebra map the generators to a 8{}-dimensional representation which 
does not need the other generators to close.  This can be easily seen 
when instead of using derivations to represent the generators, an appropriately
sized Clifford algebra is used \cite{HTT}. The use of a Clifford algebra 
may allow more insight into the whole process. This and the difference 
between using the ``small'' and the ``large'' $\cn =4$ algebras will be 
discussed in Section \ref{clifford}. 

Another point is whether the central extension should be dropped in 
the equations.  From \cite{Linch}, the closure of the algebra is found 
to be related to the existence of a central extension, specifically the 
central extension is  eliminated for $\cn >2 $. Because $\cn=4$ closes 
also, it is a valid question to ask if a central extension may exist too. 
The Jacobi Identity on $(G^{\rI}_{\ca}, U^{\rIJ}_{\cb}, G^{\rK}_{\cc})$
was used before to answer this question.  Because the supercommutators
have the same form as the $\cn >2 $, it would seem that the answer 
would be true.  But there are no longer $U^{\rIJ}_{\ca}$ generators
in the algebra.  The Jacobi identity for the other generators must be
analyzed to check if a central extension is allowed.  Although this
could be addressed now, this question will be revisited later when the
Clifford representation of the generators is presented .  For now, $c$
will be set to zero.

\setcounter{equation}{0}
\section{Description of the Coadjoint Orbit Method} \label{Coajoint}

$~~~$ A compact description of the Coadjoint Orbit method can be found
in a paper by Witten\cite{Witten} but to go into more detail and understanding, 
the work of Kirillov \cite{Kirillov} provides more insight.  To fully understand  
the Coadjoint Method, one must go to its foundation in Lie groups and 
algebras then build
from there.  A Lie group $G$  is a set of elements with certain 
topological and algebraic properties, namely continuity and
analyticity.  One can think of it as both a group of elements
and a smooth manifold.  It has a multiplication law which can be
represented as a smooth map.  The group can act on itself and 
there is a special map  defined for every point in the group: 
\begin{equation}
A_{g}(h)\equiv h\rightarrow ghg^{-1}:\forall g,h\in G
\end{equation}

Related to the Lie group is its Lie algebra $\Sf G$, a vector space 
that can be understood as the tangent space of the manifold at the 
unit point in the group, denoted by $e$ \ The unit point is a fixed
point of the previous map, meaning that it is mapped into itself.
Around this fixed  point, the derivative of $A_g$ acts to map 
elements of the Lie group to other elements in the same Lie group.  
This derivative is called the adjoint map of the Lie group:
\be
Ad_{g}:  g \rightarrow  g' \, {\rm with} \, g\,,\,g' \in G
\ee
Since elements of the Lie algebra \Sf{G} changes elements of the 
group to other elements, we find that the previous map can be 
considered a mapping of elements of the Lie algebra to other 
elements of the algebra.  Thus the map from $g$ to $Ad(g)$ can 
also be seen as
\be
Ad: (g \rightarrow Ad(g)) \simeq (\Sf{g} \rightarrow \Sf{g}'), g 
\in G, \, \Sf{g},\Sf{g}' \in \Sf{G}
\ee

This is the adjoint representation of the Lie group G.  By taking the
derivative of this map, one gets the adjoint representation of the Lie
algebra $\Sf{G}$, which has the following property:
\be
ad_{{\Sf g}} ({\Sf h}) = [{\Sf g},{\Sf h}], \, {\Sf g},{\Sf h} \in {\Sf G}
\ee
where the right{}-hand side is the Lie bracket defined for the Lie
algebra.

Since $\Sf G$ is also a vector space, we can talk about the dual
linear space ${\Sf G}^{\ast}$.  The dual space ${\Sf G}^{\ast}$ consists 
of dual elements  ${\Sf g}^{\ast}$ of  elements ${\Sf g}$ in ${\Sf G}$.  
The dual elements belong to the space of linear functions of the 
algebra element $\Sf g$.  With the definition of a bilinear form on 
both types of elements $\VEV{\sg^{\ast}, \sg}$, there also exists the 
space $\sG^{\bot}$ of elements $\sg^{\bot}$ orthogonal to the 
element $\sg$ defined by the bilinear form,  $\VEV{\sg^{\bot}, \sg}=0$.
Let P be a projection operator that projects into $\sG^{\bot}$. \ Then 
one can construct a coadjoint representation K(g) that sends elements 
of the dual space into a space of other elements orthogonal to the first:
\be
K(g) = \{ \sh \rightarrow \sg^{\bot} \equiv P({\Sf {ghg}}^{-1}) {\rm \, such 
\, that \,} \VEV{\sg^{\bot},\sg} =0, \sh \in \sG^{\ast}, \sg^{\bot} \in \sG^{
\bot}, \sg \in \sG \}.
\ee
Once one has a realization of the appropriate algebra, the
coadjoint orbit method can be applied.  First, an adjoint vector
consisting of all the generators and a central extension is
constructed.  From there, a corresponding coadjoint vector can be
formed and calculated using the ideas that
\begin{enumerate}
\item an adjoint vector can act on another adjoint vector to give 
an adjoint vector, and
\item an inner product of an adjoint vector and its dual coadjoint 
vector should be ``orthogonal'' in the sense that it gives delta functions 
in indices.
\end{enumerate}

Once the action of the adjoint vector is understood on the different
elements (which is equivalent to the first statement,) then the action of
the adjoint vector on an arbitrary coadjoint vector can be calculated.
 This now defines how the fields in the coadjoint vector transform
with respect to the elements of the adjoint vector which are related to
the underlying algebra.  The coadjoint orbit is the space of all
coadjoint vectors that can be reached by application of the action of
the algebra. 

Now the relationship of the adjoint and coadjoint elements to
sympletic structures can be utilized.  There is a relationship
between coadjoint orbits and symplectic structures.  An orbit of a 
map is like an equivalence class of the map. The coadjoint orbit 
is the equivalence class of dual linear functions on the Lie group.  
Having a symplectic structure means that there exists a closed 
non{}-degenerate, skew{}-symmetric differential 2{}-form.  This 2{}-form 
is $G${}-invariant and exists for each orbit in $g^{\ast}$. Having a 
symplectic structure is also related to Poisson brackets and phase 
space.

The infinitesimal version of the coadjoint action is
\be
\VEV{K(\sg)\sh, \sg'} = \VEV{\sh, -ad_\sg(\sg')} = \VEV{\sh, [\sg,\sg']}, \sg,\sg' \in 
\sG, \sh \in \sG^{\ast}.
\ee
This is equivalent to the natural skew symmetric bilinear form,
$\O$, found on each coadjoint orbit. The form
$\O$ is defined on adjoint elements as
\be
\O_{B}(\tilde B_1 , \tilde B_2) = \VEV{\tilde B, [a_1, a_2 ]}
\ee
with $a_{1}$ and $a_{2}$ as associated fields from the adjoint 
vector and  $\tilde {B}$ the coadjoint vector.  The change in $\tilde 
B$ from the specific adjoint fields is given by
\be
\delta _{a_{i}}\tilde {B}=a_{i}\ast \tilde {B}  .
\ee
So the inner product of a adjoint and coadjoint element also generates
the infinitesimal variation of the corresponding coadjoint field with
respect to the symmetry generated by the adjoint element.

 To get to the calculation of the Operator Product Expansions, one
more step must be done.  The physical fields and the conjugate
momentum must be associated with the  adjoint and coadjoint elements. 

As the adjoint element generates a symmetry transformation of some 
kind, associated with that transformation is a charge, $Q_{i}$.  For the 
adjoint field $a_{i}$, the charge can be calculated as
\be
 a_{i}\rightarrow Q_{a_{i}}=\int d\tau \, G^{i}a_{i}  
\ee
$G^{i}$ is the generating function for the transformation and comes from 
the action of the adjoint element on the coadjoint vector represented by 
the associated fields.

Using some concepts from mechanics, one can see that this charge
generates the infinitesimal variation of a function of a field, $f_{i}$, and 
its conjugate field, ${\pi}^{i}$, through the use of Poisson Bracket:
\be
\{Q_{a_{i}},F(f_{i},\pi ^{i})\}=\frac{\partial Q}{\partial f_{i}}\frac{\partial F}
{\partial \pi ^{i}}-\frac{\partial Q}{\partial
\pi ^{i}}\frac{\partial F}{\partial f_{i}}=-\delta _{a_{i}}F(f_{i},\pi^{i})
\ee
There are three fields in this equation: the adjoint field related to the 
transformation, $a_{i}$; a coadjoint field, $f_{i}$; and the conjugate 
momentum field to $f_{i}$, ${\pi}^{i}$.

Defining A to the be the dual coadjoint field to adjoint field
$a_{i}$ , then
\be
(a_{i},A)=a_{i}\ast A=\int a_{i}(x)A(x)\,{dx}=\,{const.}
\ee
And it can also be shown that
\be
\{Q_{a_{i}},F(x)\}=a_{i}(x)\frac{\delta }{\delta a_{i}(x)}F(x)=\int \, {dy} \, a_{i}(y) 
\,[A(y)F(x)]
\ee

The quantity in brackets on the RHS gives the short distance OPE between
A, the dual coadjoint element of ${\alpha}$, and F, a function of the phase 
space elements. By using this equation, the adjoint action on coadjoint 
elements can be mapped to the infinitesimal variation of the dual fields by 
the symmetries generated by the algebra elements.  The OPE can almost 
be read off from the resulting equation.

 The actual use of the method flows from the following steps:
\begin{enumerate}
\item Choose an coadjoint field and an adjoint action on it.  This gives the
variation of the physical field with respect to some transformation.
\item Calculate the Poisson bracket of the charge generated by the adjoint
action on the the physical field.  The generating function of the
transformation will come from the calculations of the adjoint action on
the coadjoint vector done earlier.
\item Compare to the integral form of Poisson bracket.  The short distance
OPE will be the equivalent expression of the previous step once it has
been put in the associated integral form. This will involve the use of
delta functions on the space (a line in the 1D case) and its
derivatives.
\end{enumerate}

Typically, one needs an action to determine the useful field theory quantities 
such as correlation functions.  However, these quantities are dependent on 
the symmetries found in the theory and not necessarily obvious in the action.  
The Coadjoint Orbit method allows for these quantities to be calculated 
without an action and totally based on the underlying symmetries of the theory 
being studied. 

As an aside, one of the uses of coadjoint orbits is relate the classification of 
the orbits to the classification of another related mathematical structure.  For 
example, if G is the set of all linear $n \times n$ real invertible matrices, then 
the classification of coadjoint orbits is equivalent to the classification of matrices 
up to similarity.  The analysis of the coadjoint orbits allows one to classify two 
dimensional conformal field theories
(2D{}-CFT's).

\setcounter{equation}{0}
\section{ Calculation of Short Distance Operator \\ $~$ Product
Expansions}\label{OPE}

$~~~$ The Operation Product Expansion (OPE) is an expression of the
product of two operators as a sum of singular functions of other operators.  
This is useful when calculating the product of field operators at the same 
point.  Wilson and Zimmerman \cite{Wilson} have a discussion of the use 
of OPEs in Quantum Field Theory.  In this case, the operators are tensor 
fields.  The general form of an OPE is
\be
A(y)B(x)\sim \sum
_{i}{ {C_{i}(x)} {(y-x)^{-i}}}+({\rm non \,singular \,terms})
\ee
where $C_i$ is a member of a complete set of operators.  The non{}-singular 
terms are not important because the singular terms determine the properties 
of the product of operators. 

The goal is to express the product of fields that represent the underlying 
algebra in terms of functions of other fields which represent other 
elements in the algebra. These products are further related to useful field 
theory quantities such as propagators and mass terms.

The methods used are found in \cite{Lubna, Curto, Linch, Bah}. Applying this 
process to the algebra of interest, the adjoint vector of the \cnd $\cal {GR}$ 
SVA is $L=(L_{A},G_{B}^{I},T_{C}^{\rJK})$.  The adjoint acting on this gives
\be
\begin{array}{ccc}
ad((L_{\cm},G_{\cn}^{\rK},T_{\cp}^{\rm LM}))(L_{\ca},G_{\cb}^{\rI},T_{C\cc}^{
\rJK})&=&(L_{\cm},G_{\cn}^{\rK},T_{\cp}^{\rm LM}) \ast (L_{\ca},G_{\cb}^{\rI},
T_{\cc}^{\rIJ})\\
&=&(L_{\cq,{\rm new}},G_{{\cal R},{\rm new}}^{\rm H},T_{\cs,{\rm new}}^{\rm FG})
\end{array}
\ee
The coadjoint element is $\tilde {L}=(\tilde
{L}_{\ca},{\tilde {G}}_{\cb}^{\rI},{\tilde {T}}_{\cc}^{\rJK})$ and correspondingly 
gives
\be
\begin{array}{ccc}
ad((L_{\cm}\, , \, G_{\cn}^{\rK}\, , \, T_{\cp}^{\rm LM }))\,(\tilde {L}_{\ca},{\tilde 
{G}}_{\cb}^{\rI},{\tilde {T}}_{\cc}^{{\rJK}})&=&(L_{\cm},G_{\cn}^{\rK},T_{\cp}^{{
\rm LM}}) \ast ({\tilde {L}}_{\ca},{\tilde {G}}_{\cb}^{\rI},{\tilde {T}}_{\cc}^{{\rJK}})\\
&=&({\tilde {L}}_{\cq,{\rm new}},{\tilde {G}}_{{\cal R},{\rm new}}^{\rm H},{\tilde {
T}}_{{\rm S},{\rm new}}^{{\rm FG}})
\end{array}
\ee

and the inner product is
\be
\langle ({\tilde {L}}_{M},{\tilde {G}}_{N}^{K},{\tilde
{T}}_{P}^{{LM}})|(L_{A},G_{B}^{I},T_{C}^{{JK}})\rangle
=\delta _{M,A}+\delta _{N,B}\delta _{K}^{I}+\delta _{P,C}\delta
_{{LM}}^{{JK}}
\ee

To calculate the OPEs, one needs the expression of  $\delta_{{L}}{\tilde {L}}
={L}\ast {\tilde {L}}$ where  ${L}$ is an adjoint vector and  ${\tilde {L}}$  is
a coadjoint vector. Using the fact that  $\langle {\tilde {L}}|{L}\rangle $ is an 
invariant and  ${L}\ast {\tilde {L}}$ can be calculated from  $\langle
{L}'\ast {\tilde {L}}|{L}\rangle $ , one can use the Leibnitz rule on the
invariant form and get
\be
\langle {L} \ast {\tilde{L}}| {L}\rangle =-\langle {\tilde {L}}|{L}\ast {L}\rangle 
\ee

Since  ${L}$ and  ${\tilde {L}}$ are made up of components (L, G, T), it 
is easier to calculate pairs of adjoint elements acting on coadjoint elements.  
This reduces the number of calculations greatly.  The list of adjoint/coadjoint 
pairs are
\begin{eqnarray*}
\delta \tilde {L} &=& L\ast \tilde {L}+G\ast \tilde {G}+T\ast \tilde {T}\\
\delta \tilde {G} &= & L\ast \tilde {G}+G\ast \tilde {L}+G\ast \tilde {T}+
T\ast \tilde {G}\\
\delta \tilde {T} &= & L\ast \tilde {T}+G\ast \tilde {G}+T\ast \tilde {T}
\end{eqnarray*}
This checks against the calculations from \cite{Curto}.  Note that there is
no  $\tilde {T}\ast L$ term in the list of changes to the coadjoint vector.

Using a realization of the algebra as tensor fields, the adjoint
representation elements are $F=(\eta ,\chi ^{I},t^{{\rm RS}})$ , which are 
general elements of the Virasoro, Kac{}-Moody, and so(4) algebras 
respectively. The coadjoint fields are  $B=(D,\psi ^{I},A^{{\rm RS}})$ , a 
rank two pseudo tensor, a set of 4 spin{}-3/2 fields, and the 6 so(4) gauge 
fields.

 The coadjoint action can be seen as generating the changes in the
fields.  It acts as
\be
F\ast {\tilde {B}}=\delta _{F}\tilde {B}=(\eta ,\chi
^{\rJ},t^{{\rKL}})\ast (D,\psi ^{\rI},A^{\rJK})=(\delta
D,\delta \psi ^{\rI},\delta A^{\rJK}).
\ee

There are three charges, one for each adjoint element/operator:
\begin{eqnarray}
L_{A}  \rightarrow   \eta & \rightarrow & Q_{\eta }=\int dx \, G_{a} \eta^{a} \\
G_{A}^{I} \rightarrow  \chi^{\rI} & \rightarrow & Q_{\chi ^{\rI}}=\int dx \, 
G_{a}(\chi ^{\rI})^{a} \\
T_{A}^{{IJ}}  \rightarrow  t^{{\rIJ}} & \rightarrow & Q_{t^{{\rIJ}}}=\int dx \, 
G_{a}(t^{\rIJ})^{a} 
\end{eqnarray}

Choosing  $L\ast \tilde {L}$ as an example, the physical field
representation is used :
\be
L\ast \tilde {L}\rightarrow \delta _{\eta }D
\ee
\be
L_{\eta}\ast {\tilde {L}}_{D}  \rightarrow \tilde {L}_{\tilde {D}}:\tilde {D}=-D'\eta 
-2D\eta'
\ee
\be
\delta _{\eta }D=\tilde {D}
\ee
\be
\delta _{\eta}D=-\{Q_{\eta },D\}=\int dy \, \eta (y)(D(y)D(x))
\ee
\be
Q_{\eta }=\int dx \, G^{a}\eta _{a}=\int dx(-D' \eta -2D\eta') \eta _{a}
\ee
\be
\{Q_{\eta },D\}=\int dy (-D'(x)\eta (y)-2D(x)\eta'(y)).
\ee

Using the 1D formula for the delta function,
\be
\delta (y-x)=\frac{1}{2\pi i(y-x)}
\ee
and integration by parts to separate out  $\eta (x)$ terms,
\be
\begin{array}{lcccc}
\{Q_{\eta },D\}&=&\int dy & \underbrace{(\partial_{x}D(x)\frac{-1}{2\pi i(y-
x)}+D(x)\frac{-1} {\pi i(y-x)^{2}})} & \eta
(x) \\
&=&\int dy & [D(y)D(x)] & \eta (x)
\end{array}
\ee

Thus by taking pairs of individual adjoint elements acting on individual 
coadjoint elements, the OPE's can found.

\begin{enumerate}
\item 
$D(y)O(x)$
\begin{eqnarray}
L_{\eta }\ast {\tilde {L}}_{D} = \tilde {L}_{\tilde {D}} &\rightarrow& \tilde {D}=
-D'\eta -2 D\eta'\\
L_{\eta }\ast {\tilde {G}}^{\bar{Q}}_{\psi ^{\bar{Q}}}={\tilde {G}}_{{\tilde {\psi 
}}^{\bar{Q}}}^{\bar{Q}} 
&\rightarrow& {\tilde {\psi}}^{\bar{Q}}=-(\frac{3}{2}\eta '\psi ^{\bar{Q}}-\eta 
(\psi ^{{\bar {Q}}'}))\\
L_{\eta }\ast {\tilde{T}}_{A^{\bar{I}\bar{J}}}^{\bar{R}\bar{S}}={\tilde {T}}_{\tilde {
A}^{\bar{I}\bar{J}}}^{\bar{R}\bar{S}} &\rightarrow& {\tilde {A}}^{\bar{I}\bar{J}}=-(
A^{{RS}})'\eta -\eta 'A^{{RS}}
\end{eqnarray}

These expressions yield the following OPEs:
\begin{eqnarray}
D(y)D(x) &=& \frac{-1}{\pi i(y-x)^{2}}D(x)-\frac{1}{2\pi (y-x)}\partial_{x}D(x) \\
D(y)\psi ^{Q}(x) &=&\frac{-3}{4\pi i(y-x)^{2}}\psi ^{Q}(x)-\frac{1}{2\pi
i(y-x)}\partial _{x}\psi ^{Q}(x) \\
D(y)A^{{RS}}(x) &=& \frac{-1}{2\pi i(y-x)^{2}}A^{{RS}}(x)-\frac{1}{2\pi i(y-x
)}\partial_{x}A^{{RS}}(x)
\end{eqnarray}

\item
$\psi(y)O(x)$
\begin{eqnarray}
G_{\chi ^{I}}^{I}\ast {\tilde {L}}_{D}=4i{\tilde {G}}_{\tilde {\chi}^{I}}^{I} &\rightarrow& {\tilde 
{\chi }}^{I}=-\chi ^{I}D \\
G_{\chi ^{I}}^{I}\ast \tilde {G}_{\psi^{\bar{Q}}}^{\bar{Q}} = \frac{\delta ^{I\bar{Q}}}{2}{\tilde {
L}}_{\tilde{D}}+{\tilde {T}}_{{\tilde {A}}^{I\tilde {Q}}}^{I\tilde{Q}} &\rightarrow & \tilde {D}=\left[(
\psi^{\bar{Q}})'\psi ^{I}-3(\psi
^{I})'\psi ^{Q}\right] \nonumber\\
&& {\tilde {A}}^{I\tilde{Q}} = 2(\chi ^{I}\psi ^{\tilde {Q}}-\chi ^{\tilde {Q}}\psi^{I})  \\
G_{\chi ^{I}}^{I}\ast \tilde {T}_{\tau^{\bar{R}\bar{S}}}^{\bar{R}\bar{S}}=\delta^{[{
RS}][{IQ}]}{\tilde {G}}_{\tilde {\psi}^{Q}}^{Q} &\rightarrow& \psi ^{\bar{Q}}=2(\chi^{
I})'t^{{RS}}+\chi^{I}(t^{{RS}})'
\end{eqnarray}

The OPEs are
\begin{eqnarray}
\psi ^{I}(y)D(x) &=& \frac{-3}{4\pi i(y-x)^{2}}\psi^{I}(x)-\frac{i}{4\pi (y-x)}\partial _{x
}\psi^{I}(x)\\
\psi ^{I}(y)\psi ^{Q}(x) &=& \frac{-4i}{(y-x)}\delta^{{IQ}}D(x)\\
\psi ^{A}(y)A^{{RS}}(x) &=&\frac{\pi}{i(y-x)}(\delta ^{{AR}}\delta ^{{LS}}-\delta^{{
AS}}\delta ^{{LR}})\psi ^{L}(x)
\end{eqnarray}

\item
$A(y) O(x)$
\begin{eqnarray}
T_{t^{{IJ}}}^{{IJ}}\ast \tilde {G}_{\psi^{\bar{Q}}}^{\bar{Q}}&=&2\delta ^{{QI}}G_{\tilde {
\phi}^{J}}^{J}-2\delta ^{{QJ}}G_{\tilde {\phi }^{I}}^{I} \rightarrow \psi ^{\bar{Q}}=t^{{IJ}}
\psi ^{Q}\\
T_{t^{{IJ}}}^{{IJ}}\ast {\tilde {T}}_{\tau^{\bar{R}\bar{S}}}^{\bar{R}\bar{S}} &=&-\delta^{
[\bar{R}\bar{S}]}(\delta_{{RS}}^{{JK}}+\delta_{{RS}}^{{KJ}}){\tilde{T}}_{(t^{{JK}})'\tau^{
\bar{R}\bar{S}}}^{\bar{R}\bar{S}}-{\tilde {L}}_{\tilde {D}}\delta^{[\bar{R}\bar{S}][JK]} 
\nonumber \\
 &\rightarrow & D=(t^{{JK}})'\tau ^{\bar{R}\bar{S}} 
\end{eqnarray}

Note that there is no  $T\ast \tilde {L}$ term.  However the $A^{JK}(y)D(x)$ and $A^{
JK}(y)A^{RS}(x)$ terms  are generated from the  $T\ast \tilde {T}$ action.  The OPE 
that follow are
\begin{eqnarray}
A^{{JK}}(y)D(x) &=&\frac{1}{4\pi i(y-x)^{2}}(\delta ^{{RS}}\delta ^{{JK}}-\delta^{{RK}}
\delta^{{LS}})A_{{RS}}(x)\\
A^{{AB}}(y)\psi ^{C}(X)&=&\frac{-1}{\pi i(y-x)}(\delta ^{{AC}}\psi ^{B}(x)-\delta^{{
AB}}\psi^{C}(x))\\
A^{{JK}}(y)A^{{RS}}(x)&=&\frac{1}{4\pi i(y-x)}\delta_{{AB}}^{{JKRS}}A^{{AB}}(x) 
\end{eqnarray}
\end{enumerate}

In the non{}-extended version of the algebra \cite{Lubna, Curto, Linch, Bah}, there 
are extra generators that must be added to close the algebra.  When the Coadjoint 
Orbit method is applied, these extra generators correspond to fields and have their 
own OPEs.  The fields ${\omega}$ and ${\rho}$, which correspond to the U and R 
operators respectively, have 44 and 11 independent components.  The spin of the 
fields are varied, either being 0 or $\fracm 12$ depending on the structure of the 
individual operator.  This also true for the general extended $\ell{\neq}{\pm}1$ case.  
However, the $\ell=\pm 1$ case does not have these fields or their OPEs.  Thus 
there is no difference between the regular ($\ell$=0) and extended ($\ell{\neq}0$) cases 
except when  $\ell={\pm}1$.  These cases reduce the number of operators and fields 
necessary to fully describe the theory.

\setcounter{equation}{0}
\section { Reformulation of the Coadjoint Orbit Methods $~$ Using
Clifford Algebras} \label{clifford}

$~~~$ Now a different perspective will be investigated using Clifford Algebras
instead of derivations.  Hasiewicz, Thielemans, and Troost \cite{HTT} have shown
that superconformal Lie superalgebras contain a Clifford algebra
structure in them.  By exploiting this structure, new information
can be gained by the implications of how the Clifford algebra exists in
the larger structure. 

Their method starts with a break down of the Lie superalgebra into
smaller, relevant subspaces: a Kac{}-Moody Lie algebra $KM(L)$ with a Lie
algebra $L$, a Virasoro algebra $Vir$, and subspaces $Q$ and $G$ with
underlying vector spaces respectively, $V$ and $W$. The underlying
vectors spaces of these subspaces ( $L$ for $KM(L)$, $\mathbb{R}$ for $Vir$, 
$V$ for $Q$, $W$ for $G$) are important along with a number of mappings 
that define the properties of each space.  For a fixed element $w \in W$, and
 $w,w' \in W;v,v' \in V;\Sigma ,\Sigma' \in L; a\in \mathbb{R}$,
 there are the following mappings:
\begin{eqnarray}
\[T_{m}(\Sigma ),T_{n}(\Sigma')\] &=& T_{m+n}([\Sigma ,\Sigma'])-c \,m 
K(\Sigma ,\Sigma')\delta (m+n) \\
\[L_{m},L_{n}\] &=& (m-n)L_{m+n}+(m^{3}-m)\delta(m+n)c/4 \\
\[L_{m},T_{n}(\Sigma )\]&=&-n \,T_{m+n}(\Sigma ) \\
\[L_{m},Q_{n}(v)\]&=&-(m/2+n)Q_{m+n}(v) \\
\[L_{m},G_{n}(w)\]&=&+(m/2-n)G_{m+n}(w) \\
\{Q_{m}(v),Q_{n}(v')\}&=&-b(v,v')\delta (m+n)c \\
\[T_{m}(\Sigma ),Q_{n}(v)\] &=&Q_{m+n}(R(\Sigma)v) \\
\{G_{m}(w),G_{n}(w')\}&=&2B(w,w')L_{m+n}+B(w,w')(m^{2}-1/4)\delta(m+n)c \\
&&-(m-n)T_{m+n}(\varphi(w,w')) \\
\{G_{m}(w),Q_{n}(v)\} &=&T_{m+n}(\varphi (w,v))\\
\[T_{m}(\Sigma ),G_{n}(w)\] &=&G_{m+n}(\Lambda (\Sigma)w)+m \,Q_{m+n}(
d(\Sigma ,w))
\end{eqnarray}
with $K, b, R, {\varphi}, {\Lambda},$ and $d$ all being mappings and
bilinear forms necessary to describe the superconformal Lie superalgebras.

\ There are special mappings that Hasiewicz et. al.\cite{HTT} use to associate
with the Clifford algebra:
\begin{eqnarray}
\varphi_{w}(w'):w'\in W \rightarrow \varphi (w,w')\in L\\
d_{w}(\Sigma):\Sigma \in L\rightarrow d(\Sigma ,w)\in V\\
i_{w}(a):a\in{\mathbb R}\rightarrow {\rm a w} \in W.
\end{eqnarray}

This set gives the exact series
\be
{\mathbb{R}} \stackrel{i_{w}}{\rightarrow}  W \stackrel{\varphi_{w}}{\rightarrow}  
L \stackrel{d_{w}}{\rightarrow}  V \rightarrow 0.
\ee

While this set of mappings and forms
\begin{eqnarray}
\psi _{w}:v\in V &\rightarrow &\psi (w,v)\in L\\
\Lambda _{w}:\Sigma \in L &\rightarrow &\Lambda (\Sigma)w\in W\\
B_{w}:w'\in W &\rightarrow &B(w,w')\in \mathbb{R}
\end{eqnarray}
gives the exact series
\be
0 \rightarrow V \stackrel{\psi_{w}}{\rightarrow} L \stackrel{\Lambda_{w}}{
\rightarrow} W \stackrel{B_{w}}{\rightarrow} \mathbb{R}.
\ee 

Note that all the mappings resemble adjoint actions, being based on a
fixed element $w$.

 One of the most important results is the relationship between the
dimensions of the vector spaces:
\be
 \abs{W}+\abs{V}=\abs{L}+1\label{dimvectsp}
\ee

With this relationship, one can categorize the type of algebra possible
since there are $\abs{W}$ symmetries that exist ($dim(W) = \cn$), and 
$\abs{L}$ is the dimension of the underlying Lie algebra. 

The spaces define a larger space  $S=W\oplus V\oplus L\oplus \mathbb{R}$  
of all the elements and an endomorphism ${\Gamma}$ that represents a
Clifford algebra with \ the mapping $B$ above as a definition:
\begin{eqnarray}
\Gamma _{w}(w'+v+\Sigma+a)& = &({aw}\, + \, \Lambda (\Sigma )w) \, +\, 
d(\Sigma ,w) \nonumber \\
  &+& \, (\varphi(w,w') \, +\, \psi(w,v)) \, +\, B(w,w') \\
\Gamma _{w}\Gamma_{w}'\, + \, \Gamma_{w}'\Gamma_{w} & = & 2B(w,w').
\end{eqnarray}
S is also given a metric by ${\theta}$:
\be
\theta (w+v+\Sigma+a,w'+v'+\Sigma'+a')=B(w,w')+b(v,v')-K(\Sigma,\Sigma')-{aa}'
\ee

At this point, a number of similarities to the Coadjoint Orbit method
shown earlier should be apparent. The elements of $S$ have this
particular form because the elements of all the different spaces are
now on an equal footing with each other under the Clifford algebra.
 The metric has the same form (up to some signs) as the action of the
dual element on the vectors describe in Section \ref{OPE}.

The superconformal Lie algebra is built from a vector representing the unit 
element in the space  $\mathbb{R}$ .  This element is multiplied by the basis 
elements of the Clifford algebra to get the other spaces $W$, $V$, and $L$.  
The previous mappings between spaces allows them to be separated to get 
the complicated structure needed.

The $\cn =4$ case is presented in their paper \cite{HTT} for a Clifford
algebra signature of $(0,4)$ explicitly and all other signatures by
inference.  The choice of $\ell=1$ corresponds to a 16{}-dimensional
representation of $S$ and the Clifford space.  The dimensions of the
spaces $W$, $V$, and $L$ are 4, 4, and 7 respectively as given by eq. 
\ref{dimvectsp}.  The basis vectors for W are
\be
w_{i}=\Gamma _{i}
\ee
and for V,
\be
v_{i}=\Gamma _{i}(\Gamma ^{5}-\ell)
\ee
where ${\Gamma}^{5}={\Gamma}^{1}{\Gamma}^{2}{\Gamma}^{3}{\Gamma}^{4}$,
$({\Gamma}^{2}) = 1$, and $\ell$ real, much like defined in the
derivation method. The elements of Lie algebra are given by the $\phi$
mapping with the addition of one more element:
\be
\varphi _{{ij}}=\varphi (w_{i},w_{j})=\Gamma _{i}\Gamma_{j}(i\neq j)
\ee
\be
\sigma =(\Gamma ^{5}-\ell).
\ee
The mappings and bilinear forms from above now take the form
\begin{eqnarray}
\Lambda (\varphi _{{ij}})w_{k} &=&\delta_{{jk}}w_{i}-\delta _{{ik}}w_{j}+
\ell \epsilon_{{ijkl}}w_{l}  \\
d(\varphi _{{ij}},w_{k})&=&\epsilon _{{ijkl}}v_{l}\\
\psi (w_{i},v_{j})&=&\Gamma _{i}\Gamma _{j}(\Gamma ^{5}-\ell)=-\ell
\psi_{{ij}}-\frac{1}{2}\epsilon _{{ijkl}}\varphi_{{kl}}-\delta _{{ij}}\sigma \\
d(\sigma ,w_{k})&=&v_{l}\\
R(\varphi _{{ij}})v_{k}&=&\delta _{{jk}}v_{l}-\delta_{{ik}}v_{j}+\ell \epsilon _{{ijkl}}v_{l} 
\end{eqnarray}

Now the correspondence between derivation representation and Clifford
algebra representation should be clear:
\begin{center}
\begin{tabular}{|c|c|c|}
\hline
\multicolumn{3}{|c|}{ \bf Table 2: Correspondence between GR Super 
Virasoro Algebra Generators } 
\\ \hline\hline
${\rm Original \, Notation}$ & ${\rm Condensed \, Form}$ & ${\rm from \, HTT}$ \\ \hline
$P \, , \, \Delta \, , \, K$ & L & $\theta (B,b,K)\in {Vir}$ \\ \hline
$ Q \, , \, S$ & G & $w_{i}\in W,v_{i}\in V$ \\ \hline
$T$ & T & $\varphi ,\sigma \in L$ \\ \hline
\end{tabular}
\end{center}
\vskip.2in
\centerline{{\bf Table 2}} 

The effects of the extended algebra, which is a function of $\ell$, can be
seen in the mapping $b$, the metric term for the vector space V, and the
mapping ${\Lambda}$ on the six linear combinations 
$\varphi_{{ij}}\pm \frac{1}{2}\epsilon _{{ijkl}}\varphi_{{kl}}$ :
\be
b(v_{i},v_{j})=-\delta _{{ij}}(1-\ell^{2})
\ee
\be
\Lambda (\varphi _{{ij}}\pm \epsilon _{{ijkl}}\varphi_{{kl}})w_{m}=(1\mp \ell)
(\delta _{{jm}}w_{i}-\delta_{im}w_{j}\mp \ell \epsilon_{{ijmn}}w_{n})
\ee

The parameter $\ell$ can be used to categorize all types of 1D $\cn = 4$ super
Virasoro algebras. \ When $\ell \neq \pm 1$, the algebra is the ``large'' $\cn =4$ 
algebra with $so(4)=so(3) \otimes so(3)$.  At $ \ell=\pm 1$, it collapses to the 
``small'' 8{}-dimensional $\cn =4$ algebra.  It is called small because at $\ell=\pm1$, 
part of the space is mapped into zero into W.  The dimension $\fracm 12$ fields
generated by V disappear and the corresponding representation now only has 
4 dimension{}-1/2 fields from W and four dimension{}-1 fields from the combination 
of $L$ and $Vir$. 

It is clear that the addition of the $\ell${}-terms, which also involved the Levi{}-Civita 
tensor, has its basis in the $Q$ vector space describing the supersymmetric 
operators and requires the necessary adjustments to the other operators to close 
the algebra. The original $[T,Q\}$, $[T,S\}$,and $[T,T\}$ supercommutators reflect 
this relationship and the close ties between the supersymmetric operators and the 
Lie algebra underneath.

The question of whether the algebra has central charges can now be revisited. 
In \cite{HTT}, the commutation relations, which are given earlier, contain the 
central charge $c$.   They make the assumption of a nonzero central charge 
and show that the set of generators is closed.  With some additional work, the 
central charge can be re-added into the equations.

With the algebra elements written as elements of a Clifford algebra, all of the 
previous work can be double checked and reanalyzed in a different context. 
The benefit of going to a Clifford algebra representation is that the Clifford 
algebras are well{}-known and well{}-understood. In \cite{HTT}, there is some 
discussion about what this  would entail and will be investigated for future 
research.

\setcounter{equation}{0}
\section{Discussions, Interpretations, and Conclusions}\label{conclusions}

$~~~$ There are a number of interesting ideas and directions that this work
has brought up:

\begin{itemize}
\item Coadjoint Orbit Method: The Coadjoint Orbit method has a clear mathematical 
basis underlying it. There exists a relationship between the equivalence class of
linear functions on a Lie group (trajectories) and a natural sympletic structure 
on the relevant manifold (phase space). The connection between the two seems 
more obvious in terms of Clifford algebras, which has a foot in 
both worlds. \ It may be that a simpler explanation can be found by exploring 
this direction with the first step going from the Clifford algebras to the underlying 
Spin groups and algebras which are closer related to Lie groups.
\item Higher{}-point functions (3{}-point and 4{}-point correlators):  The methods 
of this paper describe using any representation of symmetry generators to
develop OPEs describing two{}-point correlation functions. \ In \cite{Wilson}, there 
is a way to extend this methodology to higher point functions.  Thus,  it may be 
possible to totally ``skip'' Hamiltionian and Langrangian and just calculate 
correlation functions from symmetries.  Skipping 
that step, however, does not absolve one from still figuring out the dynamics of the 
theory, which are contained in the propagation and interaction terms calculated 
from the OPEs.
\item Since the Virasoro and Kac{}-Moody algebras are Lie algebras, they
have interpretations as manifolds.  What does the central extension mean 
in terms of manifolds?  A central extension in group representation terms means 
that there are operators (or combinations of operators) that exist in the center of 
the group besides the typical identity element.  The formal name for this concept 
is an ideal, a subgroup that maps products between members inside and outside 
the subgroup into the subgroup. In this case, it represents  that elements in the 
group can be pulled back to ``another origin''.  The interpretation of the central 
extension should be important for any work involving Geometrical Representation 
theory.
\item In \cite{HTT}, they discussed the non{}-existence of a description of 
superconformal Lie superalgebras with dim $W > 4$.  There were a number of 
restrictions to this statement but they discuss $\cn > 4$ superconformal superalgebras 
that were not Lie superalgebras.  Further research into this area could provide a
possible generalization of supersymmetry algebras.
\end{itemize}

Our use of super vector fields in order to realize the symmety generators in a 
geometrical manner also points in one other direction.  Since there is no metric 
defined on a Salam-Strathdee superspace, the conventional and familiar role 
of the metric (or a putative super-metric) is taken over by super-frame fields or 
super vielbeins.  Thus a definition of Killing super-vectors must rely on a super 
vielbein.  As such there is a superspace geometry that is naturally associated 
with the  vector fields (realizing the symmetry).  This geometry is the conventional 
one of a flat Salam-Strathdee superspace.  This raises a question.  One can imagine 
a super vielbein that does {\em {not}} describe a flat  Salam-Strathdee superspace
but one with a non-trivial topology.  If it possesses a related set of Killing super 
vectors.  In principle it should be possible to derive short distance expansions in 
this case.  

In conclusion, the short distance OPEs for the extended 1D $\cn =4$ Super 
Virasoro algebra was calculated and found to be exactly of the same form 
of the 1D $\cn =2$ case.  Further investigation showed the full relationship 
between the ``large'' and ``small'' $\cn =4$ algebras and the deeper relationship 
between the two through the Clifford algebra.   Let us end by noting that the relation
to Clifford algebras also suggest that `Garden Algebras' defined in \cite{GRdN}
seem likely to provide a starting point for some OPE's. \vskip 0.1in

${~~~~~}$``{\it {No human investigation can be called
real science if it cannot be }}${~~~}$ \newline
${~~~~~~~~~~~}${\it {demonstrated mathematically.}}''
\newline $~~~~~~~$ -- Leonardo da Vinci
\newline ${~~~}$

\noindent
{\Large\bf Acknowledgments}

This research has been supported in part by NSF Grant PHY-06-52363,
the J.~S. Toll Professorship endowment and the UMCP Center for
String \& Particle Theory.

\end{document}